\documentclass[epj]{svjour}
\usepackage{latexsym}
\usepackage{graphicx}
\usepackage[usenames,dvipsnames]{xcolor} 
\usepackage{amsmath,amssymb}

\begin{document}
\title{Cosmic censorship, massless fermionic test fields, and absorption probabilities}
\author{Koray D\"{u}zta\c{s}
}                     
%

\institute{Department of Natural and Mathematical Sciences,
\"{O}zye\u{g}in University, 34794 \.{I}stanbul Turkey }

\date{Received: date / Revised version: date}
%
\abstract{ In the conventional approach, fermionic test fields lead to a generic overspinning of black holes resulting in the formation of naked singularities. The absorption of the fermionic test fields with arbitrarily low frequencies is allowed  for which the contribution  to the angular momentum parameter of the space-time diverges. Recently we have suggested a more subtle treatment of the problem considering the fact that only the fraction of the test fields that is absorbed by the black hole contributes to the space-time parameters. Here, we re-consider the interaction of massless spin $(1/2)$ fields with Kerr and Kerr-Newman black holes, adapting this new approach. We show that the drastic divergence problem disappears when one incorporates the absorption probabilities. Still, there exists a range of parameters for the test fields that can lead to overspinning. We employ backreaction effects due to the self-energy of the test fields which fixes the overspinning problem for fields with relatively large amplitudes, and renders it non-generic for smaller amplitudes. This non-generic overspinning appears likely to be fixed by alternative semi-classical and quantum effects.
%
} 
\maketitle
\section{Introduction}
Penrose singularity theorem implies that a space-time fails to satisfy geodesic completeness following the formation of a trapped surface, during gravitational collapse \cite{pensing}. Geodesic incompleteness is identified with the existence of a singularity. One way to maintain the smooth causal structure of the space-time is to disable the causal contact of the singularity with distant observers. This requires the singularities to be covered by event horizons. The cosmic censorship conjecture states that the gravitational collapse should  end up as a black hole rather than a naked singularity; thus forbids the causal contact of singularities with distant observers \cite{ccc}.

By definition, a black hole is an object surrounded by an event horizon. Penrose singularity theorem does not imply that the black holes are generic solutions of general relativity unless the cosmic censorship conjecture is valid. However, a rigorous proof of the cosmic censorship conjecture has been elusive for decades. A closely related problem is the possibility to perturb a black hole by test particles and fields to destroy the event horizon. This problem was first studied by Wald~\cite{wald74}. The main concern in both problems is whether or not a space-time can include a naked singularity which can be in causal contact with distant observers. In other words we would like to test if the singularity is ``censored'', and if it remains ``censored''.

In Wald type problems one attempts to increase the angular, momentum or charge parameter of a black hole beyond the extremal limit. If this can be achieved the event horizon can be destroyed to expose the singularity. Following Wald various thought experiments were constructed to test the validity of cosmic censorship. These thought experiments involve  perturbations of the black holes with both test particles \cite{hu,js,backhu,backjs,f1,hodself,gao,siahaan,magne,yuwen,higher,v1,he,wang,gim,jamil,shay,shay2,shay3,shay4,zeng}, and fields \cite{semiz,overspin,emccc,duztas,toth,natario,duztas2,mode,taub-nut,kerrsen,gwak5,gwak6,hong,gwak1,yang,kerrmog,bai,khoda,gwak2}. There were also attempts to incorporate quantum effects \cite{q1,q2,q3,q4,q5,q6,q7}, and test the validity of cosmic censorship for asymptotically anti-de Sitter cases \cite{btz,gwak3,chen,ongyao,mtz,he2,dilat,yin}. The state of the cosmic censorship conjecture has been evaluated in a recent review \cite{ong}.

In the conventional approach developed by Wald, one starts with a black hole surrounded by an event horizon with initial parameters of mass, angular momentum, charge. These initial parameters satisfy a certain inequality which assures that the event horizon exists. For example, a Kerr-Newman black hole satisfies:
\begin{equation}
M^2-Q^2-(J^2/M^2)\geq 0
\label{knmain}
\end{equation}
Next, one perturbs this black hole with test particles or fields with energy $\delta M$, angular momentum $\delta J$, and charge $\delta Q$. In the test particle/field approximation we assume that the background geometrical structure of the space-time does not change but the space-time parameters are modified.
\begin{eqnarray}
&& M \to (M+\delta M) \nonumber \\
&& J \to (J+\delta J) \nonumber \\
&& Q \to (Q+\delta Q) \nonumber \\
\end{eqnarray}
If the modified parameters of the space-time fail to satisfy the main inequality (\ref{knmain}), we conclude that the event horizon is destroyed and the final parameters of the space-time represent a naked singularity. 

Recently, Sorce and  Wald constructed an alternative method to test the possibility to destroy the event horizon \cite{w2}. For that purpose, they define a function $f(\lambda)$ such that the event horizon is destroyed if $f(\lambda)$ becomes negative. They claim that the terms first order in $\lambda$ can make $f(\lambda)$ negative. However, the contribution of the terms that are second order in $\lambda$ make $f(\lambda)$ positive again.  In \cite{absorp} we have disputed their approach and explicitly demonstrated that their method involves order of magnitude problems when one imposes the non-controversial fact that $\delta M$ is inherently a first order quantity for test particles and fields. In particular the function $f(\lambda)$ defined by Sorce and Wald has the form
\begin{equation}
f(\lambda)\sim O(\epsilon^2) - O(\lambda \epsilon (\delta M)) + O(\lambda^2 (\delta M)^{2})
\label{manifest}
\end{equation}
where $\lambda$ and $\epsilon$ are small parameters.  Note that the leading term contributes to $f(\lambda)$ to second order. Since $\delta M$ is inherently a first order quantity, the terms that are first order in $\lambda$ contribute to $f(\lambda)$ to third order (not second), and the terms that are second order in  $\lambda$ contribute to $f(\lambda)$ to fourth order (not second). Apparently, the third order terms $O(\lambda \epsilon (\delta M))$ cannot make $f(\lambda)$ negative, and the fourth order terms $O(\lambda^2 (\delta M^{2}))$ cannot fix anything. (See \cite{absorp} for an elucidative discussion) Despite the fact that the order of magnitude errors are manifest, the Sorce-Wald method is widely accepted in black hole physics.  

In \cite{absorp} we have also suggested a new approach to Wald type problems by considering the fact that only the fraction of a test field that is absorbed by the black hole contributes to the parameters of the space-time. This fraction is determined by the absorption probability which refers to the relative fluxes of the transmitted and incident modes. For the superradiant modes of the bosonic fields the absorption probability becomes negative so that the field is reflected back with a larger amplitude. In this sense, the relative flux of the transmitted and incident modes is not an actual probability as we have argued in \cite{q7}. Still, we adapted the conventional term ``absorption probability'' for the relative fluxes  in \cite{absorp} which will be retained in this paper. After all, the relative flux for fermionic fields never becomes negative and can be regarded as a probability.

In the new approach, one considers a test field with energy $ \delta M$ at infinity. The absorption probability of the test field is denoted by $\Gamma$. Since the test field is partially absorbed by the black hole and partially reflected back to infinity, the energy absorbed by the black hole is
\begin{equation}
E_{\rm{abs}}=\Gamma  (\delta M )
\label{deltaeabs}
\end{equation}
since  the rest of the energy $E_{\rm{ref}}=(1-\Gamma)(\delta M)$ is reflected back to infinity. Therefore the test field modifies the mass parameter by an amount
\begin{equation}
M \to (M+\Gamma (\delta M ))
\end{equation}
In \cite{absorp} we showed that the incorporation of the absorption probability  fundamentally changes the results of the calculations for the validity of the main inequality (\ref{knmain}), for bosonic fields.
In the conventional method, the optimal perturbations with frequency at the superradiance limit appear to be the most challenging modes to destroy the event horizon. However the absorption probability for these modes is zero which means they are entirely reflected back to infinity. These modes do not modify the original parameters of the spacetime therefore they do not constitute a challenge for the event horizon when one takes the absorption probabilities into consideration.  We have also shown that only a small fraction of the challenging modes with frequencies close to the superradiance limit, is absorbed.  Incorporation of the absorption probabilities gives us the ultimate solution for the overspinning problem due to bosonic test fields. 

However the case is fundamentally different for fermionic fields. The energy momentum tensor for the fermionic fields does not satisfy the weak energy condition and they do not exhibit superradiant scattering. The absorption of fermionic fields with arbitrarily low frequencies is allowed. In \cite{generic} we have shown that this leads to drastic results concerning the validity of cosmic censorship. The contribution of test fields to the angular momentum parameter of the black hole is inversely proportional to the frequency $\omega$; namely $\delta J=(m/\omega)\delta M$. In the absence of a lower limit for $\omega$, $\delta J$ increases without bound which leads to a generic overspinning of black holes. Backreaction effects become irrelevant far before $\omega$ approaches to zero. Since the overspinning is generic, we postponed the solution of the problem to a quantum theory of gravity  beyond the semi-classical approximation, which does not appear to be imminent.

In this work we investigate whether a more subtle treatment of the overspinning problem which we have proposed in \cite{absorp}, can fix the overspinning problem for fermionic fields. In section (\ref{kerr}), we evaluate the interaction of Kerr black holes with massless spin $(1/2)$ fields by taking the absorption probabilities into consideration and compare the results with the conventional method. We derive that there exists a range of parameters for the test field which can lead to overspinning. However overspinning is not generic anymore and it is prone to be fixed by employing backreaction effects. In section (\ref{backkerr}) we show that the self-energy of the test fields due to the increase in the angular velocity of the event horizon fixes the overspinning problem for fields with a relatively large magnitude and renders it non-generic for small amplitudes. In section (\ref{kerr-newman}) we extend the results derived for Kerr black holes to Kerr-Newman black holes.

\section{Fermionic fields and Kerr black holes}
\label{kerr}
The well-known effect of superradiance refers to the fact that bosonic fields interacting with Kerr black holes get reflected back to infinity with a larger amplitude if the frequency of the field is lower than the critical value:
\begin{equation}
\omega_{\rm{sl}}=m \Omega=\frac{m a}{r_+^2 + a^2}
\label{srlimit}
\end{equation}
where $\omega_{\rm{sl}}$ is the superradiance limit. In other words, if the frequency of the field is below the superradiance limit no net absorption of the test field occurs. In an alternative approach, it has been shown  a test particle or a field cannot be absorbed by a Kerr black hole unless the contributions to mass and angular momentum parameters satisfy
\begin{equation}
\delta M \geq \Omega \delta J
\label{needham}
\end{equation}
The first derivation of this condition known to this author is by Needham \cite{needham}. For test fields with 
\[ \delta J =\frac{m}{\omega} \delta M
\]
Needham's condition gives identically the same result for the minimum value of the frequency of a test field to allow its absorption by a Kerr black hole
\begin{equation}
\omega \geq m\Omega
\label{lowerlimit}
\end{equation}
The relative contribution of a test field to the angular momentum parameter of the black hole is inversely proportional to its frequency $\omega$. If the absorption of test fields with arbitrarily low frequencies were allowed, their contribution to the angular momentum parameter would be much larger compared to their contribution to the mass parameter. In particular this contribution would diverge to infinity as $\omega$ approaches zero. For that reason, the existence of a lower limit to allow the absorption of a test field is crucial to prevent the overspinning of black holes into naked singularities. The overspinning would become inevitable without the existence of the lower limit (\ref{lowerlimit}).

The derivation of the superradiance effect and Needham's condition are based on the assumption that the energy momentum tensor of the test field satisfies the weak energy condition. However it is known that fermionic fields do not satisfy the weak or the null energy condition. Superradiance does not occur for fermionic fields and Needham's condition does not apply. There is no lower limit to allow the absorption of fermionic field; i.e. the absorption probability approaches zero, only as $\omega$ approaches zero \cite{page}. This leads to a generic overspinning of Kerr and Kerr-Newman black holes as we have previously discussed in some of our previous works \cite{duztas,mode},  culminating in \cite{generic}.

\begin{center}
\begin{figure}
\includegraphics[scale=0.08]{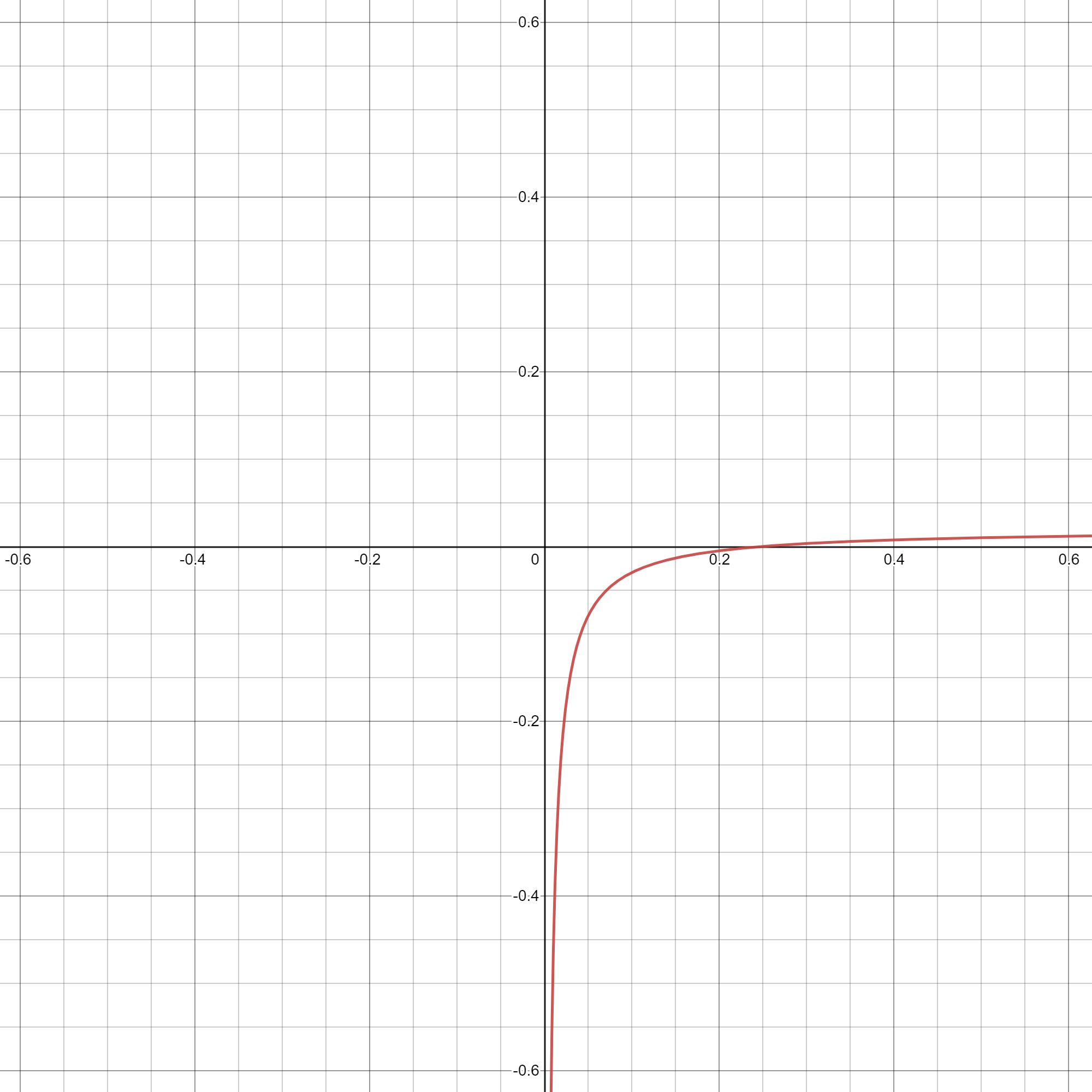}
\caption{$\Delta_{\rm{fin}}$ becomes negative around $\omega\sim 0.25$, then sharply diverges as $\omega$ approaches zero. (Here we let $M=1$, $\epsilon=\zeta=0.01$)}
\label{figure}
\end{figure}
\end{center}

In this section we send in massless spin $(1/2)$ test fields to a Kerr black hole to test the possibility of destroying the event horizon. First, we  adapt the conventional approach and show that overspinning is inevitable and generic. This can be considered as the $Q \to 0$ limit of our results for Kerr-Newman black holes in \cite{generic}. After that we  re-consider the problem by incorporating the absorption probabilities, which will fundamentally alter the course of the analysis. We start with a Kerr black hole which satisfies
\begin{equation}
M^2-J=M^2\epsilon^2
\label{kerr1}
\end{equation}
For $\epsilon \ll 1$ the black hole is nearly extremal, whereas the case $\epsilon=0$ corresponds to an extremal black hole. We send in a massless spin $(1/2)$ test field from infinity with energy $\delta M=M\zeta $ and angular momentum $\delta J =(m/\omega)\delta M$, where $m=(1/2)$ is the azimuthal wave number of the test field and $\omega$ is its frequency. In the conventional approach we assume that the final parameters of the space-time is given by:
\begin{eqnarray}
&&M_{\rm{fin}}=(M+\delta M)=M(1+\zeta) \nonumber \\
&&J_{\rm{fin}}=(J+\delta J)=J+\frac{m}{\omega}\delta M=J+\frac{m}{\omega}M \zeta
\end{eqnarray}
We define the function
\begin{eqnarray}
\Delta_{\rm{fin}}(M,J)&\equiv &M_{\rm{fin}}^2-J_{\rm{fin}} \nonumber \\
&=& M^2(\epsilon^2 + \zeta^2 + 2\zeta)-\frac{m}{\omega}M\zeta
\label{deltafinkerr1}
\end{eqnarray}
where we have imposed (\ref{kerr1}) for the initial parameters of the black hole. If the function $\Delta_{\rm{fin}}(M,J)$ becomes negative at the end of the interaction, we may conclude that the event horizon is destroyed exposing the singularity. $\Delta_{\rm{fin}}(M,J)$ will be negative if the frequency of the incoming field satisfies
\begin{equation}
\omega=\omega_{\rm{crit}} < \frac{m\zeta}{M(\epsilon^2 + \zeta^2 + 2\zeta)}
\label{omegacrit}
\end{equation}
$\Delta_{\rm{fin}}$ becomes zero for the critical value of the frequency given in (\ref{omegacrit}). If the frequency of the incident field is lower than $\omega_{\rm{crit}}$, $\Delta_{\rm{fin}}$ will be negative indicating the formation of a naked singularity. For frequencies slightly less than $\omega_{\rm{crit}}$, $\Delta_{\rm{fin}}$ will be close to zero; i.e. $\Delta_{\rm{fin}}\sim -M^2 \zeta^2$. The overspinning problem due to these modes can be fixed by backreaction effects which contribute to second order to $\Delta_{\rm{fin}}$. In our previous analysis for Kerr-MOG \cite{kerrmog}, and Kerr-Newman black holes \cite{generic} we have verified that the overspinning of Kerr black holes by scalar fields is fixed by the backreaction effects. This relies on the fact that the superradiance limit prevents the absorption of modes with frequencies much smaller than $\omega_{\rm{crit}}$. For fermionic fields superradiance does not occur; or equivalently Needham's condition (\ref{needham}) does not apply . There is no lower limit for $\omega$ to prevent the absorption of the test fields. The absorption of modes with arbitrarily low values of $\omega$ is allowed for which $\Delta_{\rm{fin}}$ diverges to minus infinity. To observe the behaviour of $\Delta_{\rm{fin}}$, we have plotted $\Delta_{\rm{fin}}$ given in equation (\ref{deltafinkerr1}) as a function of the frequency of the incoming field for $M=1$ and $\epsilon=\zeta=0.01$, in figure (\ref{figure}). $\Delta_{\rm{fin}}$ becomes zero around $\omega=0.24752$, then it sharply diverges as $\omega$ approaches zero. This divergence problem leads to drastic results considering the validity of cosmic censorship.

For a numerical example, consider two modes with frequencies $\omega_1=0.245 (1/M)$ and $\omega_2=0.1 (1/M)$. For  $M=1$, $\epsilon=\zeta=0.01$, $\Delta_{\rm{fin}}$ will be equal to 
\begin{eqnarray}
\omega_1&=&0.245 \rightarrow \Delta_{\rm{fin}} \sim -0.0002 \sim -M^2\epsilon^2 \nonumber \\
\omega_2&=&0.1 \rightarrow \Delta_{\rm{fin}} \sim -0.03
\end{eqnarray}

The overspinning due to the former mode can be fixed by backreaction effects; however the overspinning due to the latter is generic. The absorption of the latter mode is only allowed for fermionic fields which leads to a generic destruction of the event horizon. Actually, the absorption of modes with arbitrarily low frequencies is also allowed for which $\Delta_{\rm{fin}}$ diverges to minus infinity (See \cite{generic} for a general discussion involving Kerr-Newman black holes)

However, the validity of the results above are restricted to the case where one ignores the effect of absorption probabilities, which refers to the ratio of the transmitted and incident fluxes. This ratio is negative for the superradiant modes of bosonic fields. Ignoring the effect of the absorption probabilities refers to the fact that one assumes that the ratio of the transmitted and incident fluxes is 1, whenever it is positive. As we stated in the introduction we have suggested a more subtle treatment of the scattering problem in \cite{absorp}. Since the test field is partially absorbed by the black hole and partially reflected back to infinity, only the fraction of the test field that is absorbed by the black hole contributes to the mass and angular momentum parameters of the space-time. In this case the final parameters of the space-time will attain the values:
\begin{eqnarray}
M_{\rm{fin}}&=&M+ \Gamma (M\zeta) \nonumber \\
J_{\rm{fin}}&=& J+\frac{m}{\omega} \Gamma ( M \zeta)
\label{modified1}
\end{eqnarray}
where $\Gamma$ is the absorption probability of the test field. The absorption probabilies for fermionic and bosonic test fields were calculated in a seminal work by Page \cite{page}. For $s=(1/2)$ and $m=(1/2)$, the absorption probability of a test field with frequency  $\omega$ is given by:
\begin{eqnarray}
\Gamma &=& \frac{1}{4}\left(1+\frac{\Omega^2}{\kappa^2} \right)\left(\frac{A\kappa \omega}{2\pi}\right)^2 \nonumber \\
&=& M^2 \omega^2
\label{prob1}
\end{eqnarray}
where $\Omega$ is the angular velocity of the horizon, $\kappa$ is the surface gravity, $A$  is the area of the horizon. (See equation (16) in \cite{page}.) For Kerr black holes these parameters take the form
\begin{eqnarray}
&&\Omega=\frac{a}{r_+^2+a^2} \nonumber \\
&& \kappa=\frac{r_+-r_-}{2(r_+^2+a^2)} \nonumber \\
&& A=4\pi (r_+^2+a^2)
\label{kerrkappa}
\end{eqnarray}
where $r_+$ is the radius of the event horizon, and $a\equiv (J/M)$ is the angular momentum parameter.
The absorption probability $\Gamma$ is positive definite for fermionic fields which indicates that superradiance does not occur. Another interesting feature is that the absorption probability does not depend on the angular momentum of the black hole and the angular velocity of the horizon. Now, we substitute the absorption probabilities to (\ref{modified1}) and calculate the final parameters of the space-time.
\begin{eqnarray}
M_{\rm{fin}}&=&(M+  \zeta M^3 \omega^2) \nonumber \\
J_{\rm{fin}}&=& (J+  m \zeta M^3 \omega)
\label{modified2}
\end{eqnarray}
The function $\Delta (M,J)$ takes the form:
\begin{eqnarray}
\Delta_{\rm{fin}}&=& M_{\rm{fin}}^2-J_{\rm{fin}} \nonumber \\
&=& M^2+M^6 \zeta^2 \omega^4 +2M^4 \zeta \omega^2 -J -mM^3 \zeta \omega \nonumber \\
&=& M^2 \epsilon^2 +M^6 \zeta^2 \omega^4 +2M^4 \zeta \omega^2 -mM^3 \zeta \omega 
\label{deltafinabsorp}
\end{eqnarray}
The results of the thought experiment involving the interaction of  black holes with fermionic fields is fundamentally altered when one incorporates the absorption probabilities. One observes that $\Delta_{\rm{fin}}$ does not diverge to minus infinity as $\omega$ approaches to zero, contrary to its analogue derived by adapting the conventional approach in (\ref{deltafinkerr1}).  The equation (\ref{deltafinabsorp}) implies that the function $\Delta (M,J)$ re-attains its initial value as $\omega$ approaches zero.  The physical interpretation is clear. As $\omega$ approaches zero, the absorption probability also approaches zero. The test field is entirely reflected back to infinity leaving the space-time parameters invariant after the interaction. 

Still, there exists a range of parameters for the frequency of the test field which would lead to a negative value for $\Delta_{\rm{fin}}$, indicating the formation of a naked singularity at the end of the interaction. In the limiting case $\epsilon=0$ and $\zeta=0$, this range is bounded below by $\omega=0$ and bounded above by $\omega=m/(2M)=0.25 (1/M)$ which are the roots of (\ref{deltafinabsorp}) with $\epsilon=\zeta=0$. For different values of $\epsilon$ and $\zeta$, the lower bound is larger than zero, and the upper bound is smaller than $m/(2M)$.  For example, for an extremal black hole ($\epsilon=0$), perturbed by a test field with $\delta E=0.01M$ ($\zeta=0.01$) the relevant range is $0<\omega<0.24992 (1/M)$. 

Both for extremal and nearly extremal black holes (independent of the value of $\epsilon$) the function $\Delta_{\rm{fin}}$ attains its minimum value at
\begin{equation}
\omega_{\rm{crit}} \simeq  \left( \frac{m}{4M} \right)
\label{omegamin}
\end{equation}
which is valid to first order in $\zeta$.
Substituting the critical value derived in (\ref{omegamin}) to (\ref{deltafinabsorp}) we can analytically calculate the minimum value for $\Delta_{\rm{fin}}$.
\begin{equation}
\Delta_{\rm{fin-min}}=M^2 \epsilon^2 - \left(\frac{m^2 \zeta}{8}\right)M^2
\label{deltafinmin}
\end{equation}
The minimum value derived for $\Delta_{\rm{fin}}$ will be negative for relevant choices of $\epsilon$ and $\zeta$. For example, for a nearly extremal black hole with $\epsilon=0.01$ perturbed by a test field with $\zeta=0.01$  the minimum value of $\Delta_{\rm{fin}}$ can be calculated as
\begin{equation}
\Delta_{\rm{fin-min}}=-0.00021 M^2
\end{equation}
whereas for an extremal black hole perturbed by the same test field, the minimum value is
\begin{equation}
\Delta_{\rm{fin-min}}=-0.00031 M^2
\end{equation}
Since the absorption probability does not depend on the initial angular momentum parameter for fermionic fields, the calculations are identical for extremal and nearly extremal black holes except the value of $\epsilon$. The negative values for $\Delta_{\rm{fin}}$ indicate the destruction of the event horizon at the end of the interaction. However, if we choose $\delta M =0.01M (\zeta=0.01)$ for the test field, for $m=(1/2)$ (\ref{deltafinmin}) implies that
\begin{equation}
\Delta_{\rm{fin-min}} \sim - M^2 \zeta^2
\end{equation}
This suggests that the overspinning problem due to fermionic fields can be fixed by employing classical backreaction effects for $\zeta=0.01$.  For smaller values of $\zeta$, the magnitude of  $\Delta_{\rm{fin}}$ will also be small. In this case, alternative semi-classical and quantum effects can potentially fix the overspinning problem.
\section{Self-energy as a backreaction effect}
\label{backkerr}
In this section we are going to calculate the backreaction effects due to the self-energy of the test fields as they interact with Kerr black holes. The interaction of the test field with the black hole leads to an increase in the angular velocity of the horizon. In a seminal work by Will this increase in the angular velocity has been estimated as \cite{will}
\begin{equation}
\Delta \Omega =\frac{\delta J}{4M^3}
\label{will}
\end{equation}
where $\delta J$ denotes the angular momentum of the test field. In our recent works \cite{kerrmog,absorp} we argued that the induced increase in the angular velocity of the horizon leads to an increase in the superradiance limit for bosonic fields and prevents the absorption of the challenging modes. The induced increase in the angular velocity also induces a first order correction in the self-energy of the test field.
\begin{equation}
E^{(1)}_{\rm{self}}=(\Delta \Omega)(\delta J)
\label{self1}
\end{equation}
Previously it had been argued that self-energy corrections should be taken into account to check the validity of cosmic censorship \cite{hodself}. The induced self-energy contributes to the mass parameter of the space-time. If this contribution is sufficiently large, the overspinning of Kerr black holes due to the fermionic fields will be prevented. Using the expression for $\Delta \Omega $ given in (\ref{will}), the self-energy can be expressed as:
\begin{equation}
E^{(1)}_{\rm{self}}=\frac{(\delta J)^2}{4M^3}
\label{self2}
\end{equation}
This value should be added to the mass parameter to determine $M_{\rm{fin}}$. We are going to calculate the contribution of the self-energy only for the minimum value of $\Delta_{\rm{fin}}$ which was calculated in section (\ref{kerr}). If the contribution of the self-energy is sufficiently large for the minimum value, we can conclude that it is sufficiently large for all negative values. The minimum value of $\Delta_{\rm{fin}}$ was derived for a test field with zeroth order energy at infinity $\delta M=M\zeta$ and frequency $\omega\simeq 0.125 (1/M)$. In the previous section we argued that classical backreaction effects can fix the overspinning problem for $\zeta \sim 0.01$. Therefore we calculate the backreaction effects for a test field with $\zeta=0.01$. The self energy of this test field can be calculated as
\begin{equation}
E^{(1)}_{\rm{self}}=\frac{(\delta J)^2}{4M^3}=\frac{m^2 \zeta^2}{4 \omega^2 M}=0.0004M
\label{self3}
\end{equation}
This self-energy should be added to the mass parameter of the black hole. The addition of the self energy modifies the minimum value of $\Delta_{\rm{fin}}$, which becomes positive. However, we should also consider the increase in the absorption probability due to the increase in the angular velocity of the horizon. This backreaction effect works against the validity of the cosmic censorship as the absorption probability of the challenging modes increases. Though the effect of the increase in the absorption probability is small, we choose to include it in our analysis for completeness. Notice that the absorption probability (\ref{prob1}) can be written as:
\begin{equation}
\Gamma=\frac{1}{4}\left((r_+-r_-)^2+\frac{\Omega^2 A^2}{4\pi^2}\right) \omega^2
\label{prob2}
\end{equation}
For extremal black holes $(r_+=r_-)$ the modified value takes the form
\begin{equation}
\Gamma'=4\left[\left((\frac{1}{2M}+\Delta \Omega \right)^2 M^4 \right] \omega^2
\label{probmodiext}
\end{equation}
where we have substituted $\Omega=(1/2M)$ and $A=8\pi M^2$ for an extremal black hole. For the test field with with $\delta J \simeq 0.04 M^2$
\begin{equation}
\Delta \Omega=\frac{\delta J}{4M^3} \simeq 0.01 \left( \frac{1}{M} \right )
\label{deltaomega}
\end{equation}
Substituting this value in (\ref{probmodiext}), we find that
\begin{equation}
\Gamma'=1.0404 M^2 \omega^2
\label{probmodiext1}
\end{equation}
The absorption probability slightly increases due to the increase in the angular velocity of the horizon. Now, we re-calculate $M_{\rm{fin}}$ and $J_{\rm{fin}}$ for an extremal black hole which interacts with a test field with frequency $\omega=0.125 (1/M)$ and energy $\delta E=M \zeta$.
\begin{eqnarray}
M_{\rm{fin}}&=&M+\Gamma' \delta E +E^{(1)}_{\rm{self}} \nonumber \\
&=& M +1.0404M^3 \zeta \omega^2 +0.0004M
\end{eqnarray}
where we have substituted the values for $E^{(1)}_{\rm{self}}$ and $\Gamma'$, derived in (\ref{self3}) and (\ref{probmodiext1}), respectively. Similarly we can calculate  $J_{\rm{fin}}$
\begin{eqnarray}
J_{\rm{fin}}&=& J + \frac{m}{\omega} \delta M=J+\frac{m}{\omega}\Gamma' M\zeta \nonumber  \\
 &=&M^2+ 1.0404 mM^3 \zeta \omega
\end{eqnarray}
where we substituted $J=M^2$ for an extremal black hole. Now we can calculate $\Delta_{\rm{fin}}$ for an extremal black hole interacting with a test field with frequency $\omega=0.125 (1/M)$ and energy $\delta E=M \zeta$. 
\begin{equation}
\Delta_{\rm{fin}}=M_{\rm{fin}}^2-J_{\rm{fin}}=0.00047 M^2
\label{backfinminex}
\end{equation}
The positive result for $\Delta_{\rm{fin}}$ indicates that extremal black holes holes cannot be overspun by spin $(1/2)$ test fields, when one employs the backreaction effects. Had we ignored the increase in the absorption probability we would have derived a slightly larger value for $\Delta_{\rm{fin}}$; namely $\Delta_{\rm{fin}}\sim 0.00049 M^2$. The effect of the induced increase in the absorption probability appears to be small. We have chosen to include this effect in our analysis for completeness.

The self energy of the test field does not depend on the parameters of the black hole, therefore we can also use the expressions  (\ref{self3}), and (\ref{deltaomega}) for the self energy and the induced increase in the angular velocity for nearly extremal black holes. We can calculate the modified value of the absorption probability by using
\begin{equation}
\Gamma'=\frac{1}{4}\left((r_+-r_-)^2+\frac{(\Omega+\Delta \Omega)^2 A^2}{4\pi^2}\right) \omega^2
\label{probmodinext}
\end{equation}
For a nearly extremal black hole parametrised as (\ref{kerr1}), we substitute $M^2-a^2 \simeq 2M^2\epsilon^2$, which leads to:
\begin{equation}
\Gamma'=1.0410 M^2 \omega^2
\label{probmodinext1}
\end{equation}
The final parameters of the black hole are given by
\begin{eqnarray}
M_{\rm{fin}}&=&M+\Gamma' \delta E +E^{(1)}_{\rm{self}} \nonumber \\
&=& M +1.0410M^3 \zeta \omega^2 +0.0004M
\end{eqnarray}
and
\begin{eqnarray}
J_{\rm{fin}}&=& J + \frac{m}{\omega} \delta M=J+\frac{m}{\omega}\Gamma' M\zeta \nonumber  \\
 &=&M^2(1-\epsilon^2)+ 1.0410 mM^3 \zeta \omega
\end{eqnarray}
For nearly extremal black holes, we calculate the final value of the function $\Delta(M,J)$
\begin{equation}
\Delta_{\rm{fin}}=M_{\rm{fin}}^2-J_{\rm{fin}}=0.00057 M^2
\label{backfinminnex}
\end{equation}
The positiveness of the final value of the function $\Delta(M,J)$ implies that the formation of naked singularities is also prevented in the case of nearly extremal black holes. 

However the positive values derived for $\Delta_{\rm{fin}}$ in (\ref{backfinminex}) and (\ref{backfinminnex}) are only valid for test fields with $\delta M =0.01M (\zeta=0.01)$. For smaller values of $\zeta$ the self-energy which depends on $\zeta^2$, will not be sufficiently large to make  $\Delta_{\rm{fin}}$ positive. (See Equation \ref{self3}) In this case, extremal black holes (and nearly extremal black holes that are sufficiently close to extremality) can be overspun by fermionic test fields. For a numerical example if we perturb an extremal black hole with a test field with $\zeta=0.001$ and repeat the same calculation including the effect of self-energy, the minimum value of $\Delta_{\rm{fin}}$ can be calculated as
\begin{equation}
\Delta_{\rm{fin}}=M_{\rm{fin}}^2-J_{\rm{fin}}=-0.00002 M^2
\end{equation}
Though the final value of the $\Delta$ function is negative, the fact that it has a small magnitude suggests that overspinning is likely to be fixed by alternative semi-classical and quantum effects. In particular we have previously argued that the evaporation of black holes acts as a cosmic censor as it carries away the angular momentum of black holes more than their masses \cite{duztas2}. For fermionic fields  with very small amplitudes the evaporation of the black holes will dominate the effect of test fields and overspinning will be prevented. In any case, we can conclude that the overspinning of black holes by fermionic fields cannot be considered generic, when one incorporates the absorption probabilities.
\section{Ferminoic fields and Kerr-Newman black holes}
\label{kerr-newman}
Previously we have shown that Kerr-Newman black holes can be generically overspun by fermionic test fields \cite{generic}. As in the case of Kerr black holes, the generic overspinning is due to the fact that the absorption of low frequency modes are allowed. In this section we re-evaluate the interaction of Kerr-Newman black holes with neutral spin $(1/2)$ test fields by taking the absorption probabilities into consideration. The expressions for the angular velocity of the event horizon, the surface gravity, and the area of the event horizon of Kerr black holes given in (\ref{kerrkappa}) are identically valid for Kerr-Newman black holes. The expression for the absorption probability is also the same as far as neutral fields are concerned. However the radius of the event horizon is modified
\begin{equation}
r_{\pm}=M \pm \sqrt{M^2-a^2-Q^2}
\end{equation}
This leads to the modification of the absorption probability:
\begin{equation}
\Gamma=(M^2-Q^2)\omega^2
\label{probkn}
\end{equation} 
For Kerr-Newman black holes, the absorption probability is lower than that of a Kerr black hole with the same mass. A smaller fraction of the challenging modes will be absorbed by the Kerr-Newman black hole. Apparently it is less probable to overspin a Kerr-Newman black hole. In sections (\ref{kerr}) and (\ref{backkerr}) we have shown that the overspinning Kerr black holes by fermionic fields is not generic when one incorporates the absorption probabilities and employs backreaction effects. The method we have adapted  for Kerr black holes can be exploited to derive the same result for Kerr-Newman black holes. We start with a Kerr-Newman black holes which satisfies
\begin{equation}
M^2-\frac{J^2}{M^2}-Q^2=M^2 \epsilon^2
\label{paramkn}
\end{equation}
We send in a test field with energy $E=M\zeta$ and frequency $\omega$. As in the case of Kerr black holes, the test field is partially absorbed, and partially reflected back to infinity. In the final case the background parameters of the space-time are given by
\begin{eqnarray}
&&M_{\rm{fin}}=M+ \Gamma (M\zeta)=M+(M^2-Q^2)\omega^2 M\zeta \nonumber \\
&&J_{\rm{fin}}=J+\frac{m}{\omega} \Gamma ( M \zeta)=J+(M^2-Q^2)m\omega M\zeta \nonumber \\
&&Q_{\rm{fin}}=Q
\end{eqnarray}
We are going to  calculate the final value of the $\Delta$ function for an extremal Kerr-Newman black hole $(\epsilon=0)$.
\begin{equation}
\Delta_{\rm{fin}}=M^2_{\rm{fin}}-\frac{J_{\rm{fin}}^2}{M^2_{\rm{fin}}}-Q^2_{\rm{fin}}
\end{equation}
Note that for an extremal black hole $(M^2-Q^2)=(J^2/M^2)$. The final parameters of the space-time can be expressed in the form:
\begin{eqnarray}
&&M_{\rm{fin}}^2=M^2 \left(1+\omega^2 \zeta \frac{J^2}{M^2} \right)^2 \nonumber \\
&&J_{\rm{fin}}^2=J^2\left(1+\omega \zeta m \frac{J}{M} \right)^2\nonumber \\
&&Q_{\rm{fin}}^2=Q^2
\label{paramknfin}
\end{eqnarray}
First we should note that the $\Delta$ function does not diverge to minus infinity as $\omega$ approaches zero, which would have been the case if we had ignored the effect of absorption probabilities. As in the case of Kerr black holes, the $\Delta$ function re-attains its initial value after the interaction with a fermionic test field as $\omega$ approaches zero. Again, we are interested in the critical value of the frequency $\omega$ for which $\Delta_{\rm{fin}}$ attains its minimum value. By expanding $(1)/(M_{\rm{fin}}^2)$ to first order in $\zeta$ and taking the derivative of $\Delta_{\rm{fin}}$ with respect to $\omega$, we can calculate the critical value of the frequency which makes $\Delta_{\rm{fin}}$ minimum.
\begin{equation}
\omega_{\rm{crit}}\simeq \frac{m(JM)}{2(M^4+J^2)}
\end{equation}
The critical value explicitly depends on the angular momentum of the Kerr-Newman black holes. Extremal black holes with different values of angular momentum behave differently in the interaction with fermionic fields. Ignoring the backreaction effects, the minimum value of the $\Delta$ function is negative at the end of the interaction for any value of the angular momentum parameter $J$. 

For a numerical example we can start with an extremal Kerr-Newman black hole with initial parameters: $J^2/M^2=0.5$ and $Q^2=0.5$. Let us perturb this black hole with a spin $1/2$ field with energy $\delta M=0.01 M (\zeta=0.01)$ and frequency:
\[ \omega=\omega_{\rm{crit}}\simeq 0.11785 \frac{1}{M}
\]
Note that the absorption probability for this field is $\Gamma=0.5 \omega^2$. After the interaction of this field with the exremal Kerr-Newman black hole, the final parameters of the space-time will attain the values formulated  in (\ref{paramkn}).  We can calculate the final value of the $\Delta$ function.
\begin{equation}
\Delta_{\rm{fin}}=M^2_{\rm{fin}}-\frac{J_{\rm{fin}}^2}{M^2_{\rm{fin}}}-Q^2_{\rm{fin}}=-0.00020M^2
\end{equation}
The minus sign indicates that the final parameters of the space-time represent a naked singularity rather than a black hole. For larger values of $J$, the absorption probability will be larger and the final value of the $\Delta$ function will be smaller. 

In the previous section we have shown that the effect of self-energy can fix the overspinning problem for Kerr black holes for $\zeta \sim 0.01$. The expression derived for the self-energy for Kerr black holes is identically valid for Kerr-Newman black holes interacting with neutral test fields. For the test field with frequency $\omega=0.11785(1/M)$ we derive that:
\begin{equation}
E^{(1)}_{\rm{self}}=\frac{m^2 \zeta^2}{4 \omega^2 M}=0.00045M
\label{selfkn}
\end{equation}
Adding this term to the final value of the mass parameter modifies the $\Delta$ function to fix the overspinning problem. For $J^2/M^2=0.5$ we derive that $\Delta_{\rm{fin}}$ is modified as (we ignore the increase in the absorption probability)
\begin{equation}
\Delta_{\rm{fin}}=0.00114M^2
\end{equation}
In the limit $J \to 1$ the modified value of the $\Delta$ function is still positive: $\Delta_{\rm{fin}} \sim 0.001 M^2$.  Employing the backreaction effects we can conclude that Kerr-Newman black holes cannot be overspun by fermionic test fields with energy $\delta M=0.01 M (\zeta=0.01)$.

For smaller values of $\zeta$, our arguments for the Kerr case are also valid for Kerr-Newman black holes. The the effect of self-energy --which depends on $\zeta^2$-- is not sufficient to fix the overspinning problem. For example, for $\zeta=0.001$ the self energy becomes:
\begin{equation}
E^{(1)}_{\rm{self}}=\frac{m^2 \zeta^2}{4 \omega^2 M}=4.5 \times 10^{-6}M
\label{selfkn2}
\end{equation}
This self-energy is not large enough to make $\Delta_{\rm{fin}}$ positive. If we re-evaluate the previous example ($J^2/M^2=0.5$) with $\zeta=0.001$, the self-energy given in (\ref{selfkn2}) modifies the final value of the $\Delta$ function as:
\begin{equation}
\Delta_{\rm{fin}}=-7 \times 10^{-6}M^2
\end{equation}
For larger values of $J$ the final value of the $\Delta$ function will be slightly smaller. Though the negative value of $\Delta_{\rm{fin}}$ indicates the formation of a naked singularity, the fact that its magnitude is small implies that the overspinning is not generic,  as we have argued for the Kerr case. In fact for Kerr-Newman black holes the absorption probability is smaller, and overspinning is less probable. 
\section{Summary and conclusions}
Previously we had shown that fermionic test fields lead to a generic overspinning of Kerr and Kerr-Newman black holes \cite{generic}. The absence of a lower limit to allow the absorption of the test fields leads to the possibility of the absorption of the test fields with arbitrarily low frequencies. For these fields the contribution to the angular momentum parameter of the space-time diverges. This renders the backreaction effects irrelevant and the destruction of the event horizon becomes inevitable. From this point of view, a solution of the  overspinning problem in the context of classical general relativity or a semi-classical framework did not seem plausible. 

In a very recent work we have suggested a new approach to thought experiments to test the validity of the cosmic censorship conjecture. We argued that only the fraction of the test fields that is absorbed by the black holes contribute to the background parameters of the space-time \cite{absorp}. We have shown that this fixes the overspinning problem due to bosonic test fields. Here, we have adapted this new approach involving the effect of absorption probabilities, to analyse the interaction of Kerr and Kerr-Newman black holes with fermionic test fields. In section (\ref{kerr}) we have analysed the problem for Kerr black holes using both the conventional and the new approach, which allows us to compare the two approaches. We used the absorption probabilities for fermionic fields which was derived by Page \cite{page}. We showed that the results are fundamentally altered when one incorporates the absorption probabilities. As the frequency of the incident field approaches zero, its contribution to mass and angular momentum parameters of the space-time also approaches zero. (See equation (\ref{deltafinabsorp})). This is due to the fact that the test field is entirely reflected back to infinity as the absorption probability approaches zero. (The same argument also applies to bosonic fields as the frequency approaches the superradiance limit \cite{absorp}.) Still there exists a range of parameters for the frequency of the incident field that can lead to overspinning. In section (\ref{backkerr}) we considered the backreaction effects due to the self energy of the test fields. To calculate the self energy,  we used the increase in the angular velocity  of the event horizon estimated by Will \cite{will}. We also considered the increase in the absorption probability due to the increase in the angular velocity of the horizon. We calculated the backreaction effects for the minimum value of the $\Delta$ function defined in (\ref{deltafinkerr1}) and showed that the minimum value becomes positive for test fields with a relatively large magnitude; i.e. $\delta M \sim 0.01M$. For smaller amplitudes the self-energy is also small, and the final value of the  $\Delta$ function remains negative. However, the magnitude of the $\Delta$ function is very small, which suggests that the over-spinning problem can be fixed by alternative classical and quantum effects.

In section (\ref{kerr-newman}) we extended the results for Kerr black holes to Kerr-Newman black holes. We derived that the absorption probability is lower for Kerr-Newman black holes compared to a Kerr black hole with the same mass. Using this absorption probability, we showed that the final value of the $\Delta$ function does not diverge to minus infinity as $\omega$ approaches zero, however it can be negative for a judicious choice of frequency. The expression for the self energy is the same for neutral fields. We modified the final value of the $\Delta$ function and derived the same results for Kerr black holes. 

The problem of the generic overspinning due to fermionic fields appears very challenging when one ignores the effect of absorption probabilities. (Ignoring the effect of absorption probabilities corresponds to assuming $\Gamma \sim 1$) One derives that the  $\Delta$ function diverges to minus infinity. ( See equation (\ref{deltafinkerr1}) and figure (\ref{figure})•) By incorporating the effect of absorption probabilities, the drastic divergence problem is solved. When one also employs backreaction effects, one finds that overspinning is prevented for test fields with relatively large magnitudes $\zeta \sim 0.01$, and it becomes non-generic for smaller amplitudes $\zeta \leq 0.001$. At this point, it seems very plausible that alternative semi-classical and quantum effects can be employed to fix the overspinning problem for smaller amplitudes.

%
%

\end{document}